\documentclass[12pt]{iopart}

\usepackage[pdftex]{graphicx}

\begin{document}

\title{Energy-resolved neutron imaging with high spatial resolution using a superconducting delay-line kinetic inductance detector}

\author{Yuki Iizawa$^1$, Hiroaki Shishido$^{1,2}$, Kazuma Nishimura$^1$, The Dang Vu$^3$, Kenji M. Kojima$^{4,5}$ Tomio Koyama$^5$, Kenichi Oikawa$^3$, Masahide Harada$^3$, Shigeyuki Miyajima$^6$, Mutsuo Hidaka$^7$, Takayuki Oku$^3$, Kazuhiko Soyama$^3$, Kazuya Aizawa$^3$, Soh Y. Suzuki$^8$ and Takekazu Ishida$^{1,2,5}$}

\address{$^1$Department of Physics and Electronics, Graduate School of Engineering,Osaka Prefecture University, Sakai, Osaka 599-8531, Japan\\
$^2$NanoSquare Research Institute, Osaka Prefecture University, Sakai, Osaka 599-8570, Japan\\
$^3$Materials and Life Science Division, J-PARC Center, Japan Atomic Energy Agency, Tokai, Ibaraki 319-1195, Japan\\
$^4$Centre for Molecular and Materials Science, TRIUMF and Stewart Blusson Quantum Matter Institute, University of British Columbia, Vancouver, BC, V6T 2A3 and V6T 1Z4, Canada\\
$^5$Division of Quantum and Radiation Engineering, Osaka Prefecture University, Sakai, Osaka 599-8570, Japan\\
$^6$Advanced ICT Research Institute, National Institute of Information and Communications Technology, 588-2 Iwaoka, Nishi-ku, Kobe, Hyogo 651-2492, Japan\\
$^7$Advanced Industrial Science and Technology, Tsukuba, Ibaraki 305-8568, Japan\\
$^8$Computing Research Center, Applied Research Laboratory, High Energy Accelerator Research Organization (KEK), Tsukuba, Ibaraki 305-0801, Japan}
\ead{shishido@pe.osakafu-u.ac.jp}
\vspace{10pt}
\begin{indented}
\item[]October 2019
\end{indented}

\begin{abstract}
Neutron imaging is one of the key technologies for non-destructive transmission testing.
Recent progress in the development of intensive neutron sources allows us to perform energy-resolved neutron imaging with high spatial resolution. 
Substantial efforts have been devoted to developing a high spatial and temporal resolution neutron imager. 
We have been developing a neutron imager aiming at conducting high spatial and temporal resolution imaging based on a delay-line neutron detector, called the current-biased kinetic-inductance detector, with a conversion layer $^{10}$B.
The detector allowed us to obtain a neutron transmission image with four signal readout lines.
Herein, we expanded the sensor active area, and improved the spatial resolution of the detector. 
We examined the capability of high spatial resolution neutron imaging over the sensor active area of 15\,$\times$\,15\,mm$^2$ for various samples, including biological and metal ones.
We also demonstrated an energy-resolved neutron image in which stainless-steel specimens were discriminating of other specimens with the aid of the Bragg edge transmission.

\end{abstract}

%
\vspace{2pc}
%
\submitto{\SUST}
%
%
%
\maketitle

\section{Introduction}
Superconducting detectors have been recognized as one of the most successful superconducting applications. 
Generally, they have the advantages of high sensitivity, fast response, and high energy resolution, and have successfully been applied to detect cosmic rays \cite{TES, MKID_1, MKID_2} and single photons \cite{SNSPD}. 
The three most successful examples include a transition edge sensor (TES) \cite{TES}, a superconducting nanowire single photon detector (SNSPD) \cite{SNSPD}, and a microwave kinetic conductance detector (MKID) \cite{MKID_1,MKID_2}. 
TES is a kind of bolometer, in which a sharp superconducting transition edge is utilized as an ultra-sensitive thermometer \cite{TES}. 
Therefore, the sensor temperature should be kept just at a midpoint of the superconducting transition temperature $T_{\rm c}$.
The main part of the SNSPD is the meander line of the superconducting nanowire, in which a bias current is fed just below the critical current.
A tiny fraction of the nanowire itself becomes a normal state when a Cooper pair breaks by an incident photon; thus, the photon incident event is detected by probing the superconducting-normal conducting transition.
In the case of the MKID, the kinetic inductance change induced by Cooper pair breaking in a superconducting resonator is detected as a change of the resonance frequency.

Some efforts were recently made to detect a neutron beam using a superconducting detector.
A TES with a B neutron absorption layer was proposed to be used as a neutron detector \cite{TES_B}.
Two superconducting tunnel junctions (STJs) on a single crystal of Li$_2$B$_4$O$_7$ with a neutron-converter layer $^6$Li or $^{10}$B were also applied to detect neutrons \cite{STJ_1,STJ_2}.
We have been developing a unique superconducting neutron detector, called the current-biased kinetic inductance detector (CBKID), aiming at a neutron imager with higher spatial and energy resolutions via the time-of-flight (TOF) method (see Section~3.2 for details) using pulsed neutron sources \cite{MgB2_1,MgB2_2,CB-KID_1,CB-KID_2,CB-KID_3,CB-KID_4, Iizawa}.
The CBKID has superconducting meander microstrip lines, to which a finite DC bias current is fed, and a $^{10}$B neutron conversion layer.
The nuclear reaction between $^{10}$B and a neutron locally breaks the finite number of Cooper pairs in the superconducting microstrip lines (see Section~3.1 for details).
A transient change of the Cooper pair density under the DC bias current generates voltage pulses proportional to the time derivative of the local kinetic inductance.
Therefore, the CBKID can operate in a wide region of the superconducting phase.
High-resolution imaging in two dimensions with four signal readout lines was recently achieved using a combination of CBKID and delay-line method (delay-line CBKID).
A successful neutron imaging with a spatial resolution of 22\,$\mu$m \cite{CB-KID_4} was demonstrated.
Various approaches other than superconducting detectors have been used to develop a neutron imaging system with high spatial and temporal resolutions.
A gadolinium-gallium-garnet (GGG) scintillator-based neutron detector with a cooled charge-coupled device (CCD) camera has reached a spatial resolution of 11 $\mu$m with a field of view of 6$\times$6\,mm$^2$ by optical magnification using cold neutrons at the reactor source \cite{GGG_scintillator}. 
However, the method using the CCD is not suitable for highly energy-resolved imaging using pulsed neutron sources.
The highest spatial resolution of 2 $\mu$m was recently achieved using a gadolinium-oxysulfide scintillator and a cooled complementary metal-oxide semiconductor (CMOS) camera from Andor Technology with a reactor neutron source \cite{2um}.
The detector magnifies the scintillation light from the scintillator.
One must calculate the center of mass of the scintillation events to obtain a high spatial resolution of 2 $\mu$m.
Additionally, a neutron imager with a high spatial resolution and a high energy resolution can be realized using a microchannel plate (MCP) \cite{MCPs}.
A $^{10}$B-doped MCP combined with a Timepix readout \cite{Timepix} as a neutron imager was developed \cite{A.S.Tremsin_2011, A.S.Tremsin_2015}.
The nuclear reaction $^{10}$B(n, $^4$He)$^7$Li, which mainly emits a 0.84\,MeV $^7$Li ion and a 1.47\,MeV $^4$He ion, is converted into pulsed electrons, which are amplified in the $^{10}$B-doped MCP.
The signal is further electronically amplified, read out by the Timepix sensor array, and signal-processed in the FPGA circuit, thereby achieving a high spatial resolution of 55\,$\mu$m with a high time resolution of 10\,ns.
A new chip, called Timepix 3, is thought to overcome the previous limitation of the count rate in modes with a high temporal resolution \cite{Timepix3}.\par

In the wavelength dependence of the neutron transmission, the characteristic sawtooth structures, which are called the Bragg edges, appear at wavelengths where the Bragg conditions are satisfied because of a transient change of the coherent scattering.
Therefore, one can distinguish the crystal structure and crystalline quality of the samples from the Bragg edges.
The analysis of the Bragg edges gives unique information of the samples, and is an important technique in material sciences \cite{Bragg, nuetron_diff_2018}.
A pulsed neutron source, which generates neutrons with a wide energy range, is suitable for observing the Bragg edge.
Thus, a combination of the high temporal resolution of a neutron detector and TOF technique in a pulsed neutron source is promising for the Bragg edge analysis of the neutron transmission.
The delay-line CBKID has a potential for application in the Bragg edge method with a high spatial resolution.

The present work expands an active area of CBKIDs from 10\,$\times$\,10\,mm$^2$ in our preceding work\cite{CB-KID_2} to 15 $\times$ 15\,mm$^2$; consequently, the spatial resolution is improved as discussed in Section \ref{Sec.imaging}.
We demonstrate herein a clear imaging of various test specimens, including biological and metal samples over the whole sensor active area.
A successful observation of a stainless-steel Bragg edge in the neutron transmission is also discussed.

\section{Current-biased kinetic inductance detector}\label{Sec.imaging}

Detailed principles of the delay-line CBKID were described in Ref.~\cite{CB-KID_4}.
Here we briefly discussed principles for signal generation and propagation, and imaging by using the delay-line method.

The kinetic inductance in the superconductor is inversely proportional to the Cooper pair density $n_{\rm s}$. 
The transient change of $n_{\rm s}$ on the superconducting wire locally occurs at a hot spot induced by a collision of the charged particle created via the nuclear reaction between $^{10}$B and a neutron.
When a DC bias current $I_{\rm b}$ is fed into the superconducting wire, a pair of voltage pulses is generated at a hot spot within a tiny segment of the superconducting wire over the length $\Delta l$, and each pulse propagates toward both ends of the wire with opposite polarities as electromagnetic waves.
A voltage $V$ across the hot spot is expressed as follows:
\begin{equation}
\label{eq:V}
V=I_{\rm b}\frac{{\rm d}L}{{\rm d}t}\simeq I_{\rm b}\frac{{\rm d}L_k}{{\rm d}t}=I_{\rm b}\frac{\rm d}{{\rm d}t} \left(\frac{m_{\rm s}\Delta l}{n_{\rm s}q^2_{\rm s}S}\right)=-\frac{m_{\rm s}\Delta l I_{\rm b}}{n_{\rm s}^2 q_{\rm s}^2 S}\frac{{\rm d}n_{\rm s}}{{\rm d}t}, 
\end{equation}
where $S$ is the cross-sectional area of the superconducting wire; $m_{\rm s}$ and $q_{\rm s}$ are the effective mass and electric charge of the Cooper pair, respectively.
We stress that $V$ is not only the function of $n_{\rm s}$ but also that of ${\rm d}n_{\rm s}/{\rm d}t$.
This is a crucial difference with the MKID.
Because of ultra fast quasi-particle excitation, ${\rm d}n_{\rm s}/{\rm d}t$ can be sufficiently large, and thus $V$ becomes finite even if the superconducting wire remains in the superconducting zero-resistance state at a hot spot.   
It is in sharp contrast with TES and SNSPD.

\begin{figure}
\begin{center}
\includegraphics[scale=0.5]{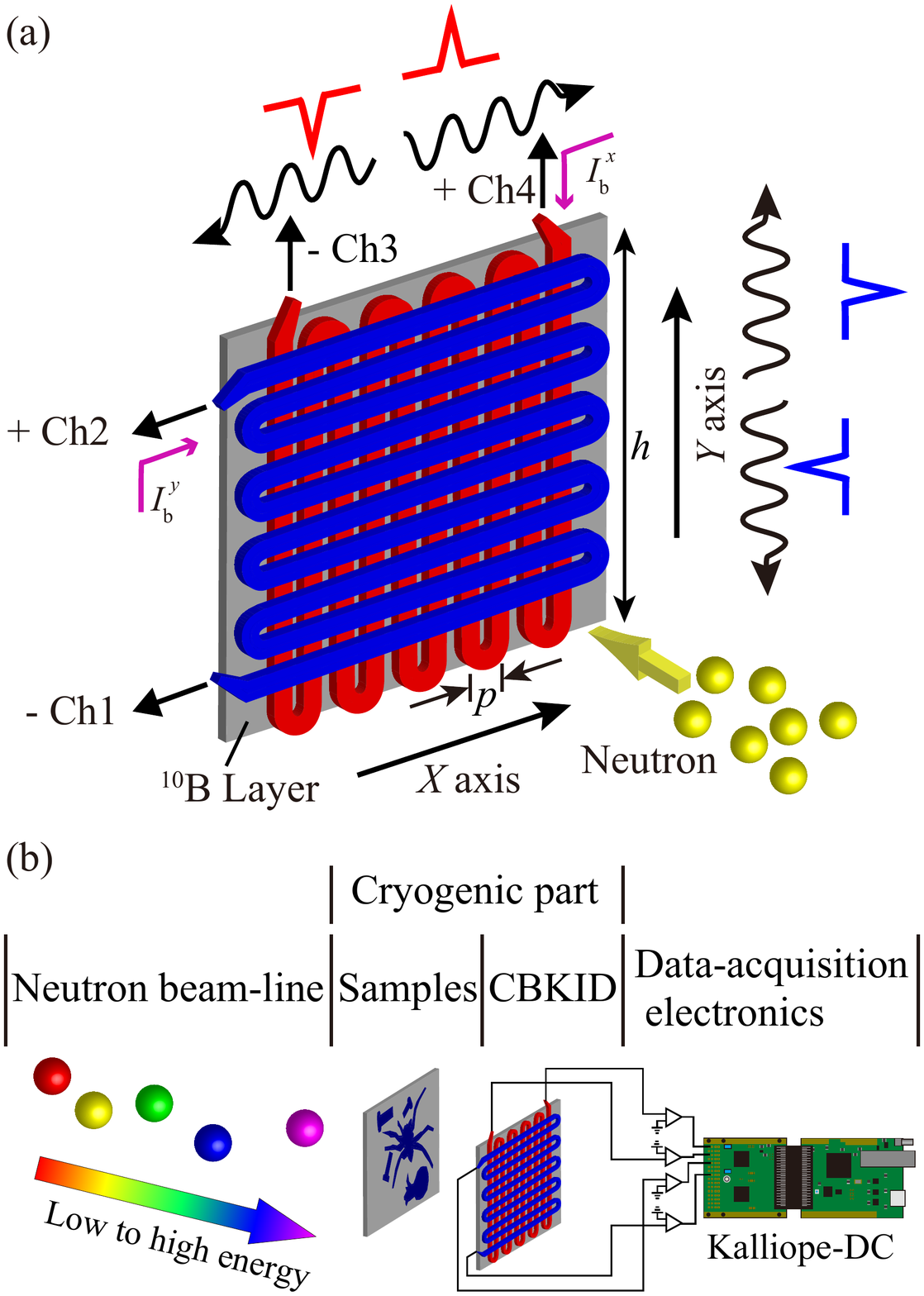}
\end{center}
\caption{Schematic of the (a) CBKID system and (b) experimental apparatus.
The $X$ and $Y$-meander lines are orthogonal to each other.
The $^{10}$B-conversion layer is deposited on the two meander lines.
Bias currents $I_{\rm b}^x$ and $I_{\rm b}^y$ are independently fed to the $X$ and $Y$ meander lines by constant DC voltage sources through 3 k$\Omega$ resistors.
Both ends of the two meander lines are connected to each channel (i.e., Ch1, Ch2, Ch3, and Ch4) of a Kalliope-DC readout circuit and an oscilloscope using four signal splitters.
Neutrons are incident from the substrate side of the CBKID.
The pulsed neutrons were irradiated to the detector through the test samples after traveling the 14-m length neutron beam-line at BL10 of the MLF of J-PARC.
CBKID and test samples were cooled down to cryogenic temperatures below $T_{\rm c}$ using a Gifford-McMahon cryocooler.}
\label{CBKID}
\end{figure}

The delay-line CBKID can image the hot spot distribution on the detector.
Figure~\ref{CBKID}~(a) shows a schematic of the CBKID system.
The CBKID has two mutually orthogonal meander lines of the superconducting Nb nanowires on the superconducting Nb ground plane.
Therefore, one regards the meander lines with the ground plane as the superconductor-insulator-superconductor (S-I-S) coplanar waveguides and expects a lower attenuation of the high-frequency traveling waves \cite{Swirhart}.
Therefore, one can observe the signals that travel through a 151\,m-length superconducting waveguide.
The signal propagation velocity can be suppressed by placing the superconducting meander line closer to the ground plane \cite{Koyama}. 
Therefore, the propagation velocities for orthogonal meander lines are different from each other.  
A more detailed discussion of the signal generation and transmission of the CBKID based on a superconducting electromagnetism has been reported in Ref. \cite{Koyama}.
As mentioned above, a pair of voltage pulses with opposite polarities appears at a hot spot on the meander line and propagates as electromagnetic waves toward both ends along the Nb meander line.

We can identify the signal quartet originating from a single event, although several signals are simultaneously present on the meander lines, as discussed elsewhere~\cite{CB-KID_4}.
Therefore, the CBKID has a high multi-hit tolerance up to the temporal resolution limit, where the signals can be discriminated.
The neutron incident positions $X$ and $Y$ are determined as follows:

\begin{eqnarray}
\label{eq:X}
X={\rm ceil}\left[\frac{(t_{\rm Ch4}-t_{\rm Ch3})v_x}{2h}\right] p,
\\
\label{eq:Y}
Y={\rm ceil}\left[\frac{(t_{\rm Ch2}-t_{\rm Ch1})v_y}{2h}\right] p,
\end{eqnarray}
where $h$ is the length of each segment of the meander line, $p$ is a repetition pitch for the meander line, $t_{\rm Ch1}$, $t_{\rm Ch2}$, $t_{\rm Ch3}$, and $t_{\rm Ch4}$ are the corresponding time stamps of the signals received at Ch1, Ch2, Ch3, and Ch4, which correspond to the signals propagated toward both ends of the $X$ (Ch3, Ch4) and $Y$ (Ch1, Ch2) meander lines \cite{CB-KID_4}.
Similarly, the $Y$ position can also be determined; hence, we can image the positions of the mesoscopic excitations in a two-dimensional (2D) plane using a very limited number (four) of electric leads for the readout circuits. 
We note that the pixel size is proportional to $v_{x, y}$ and inversely proportional to $h$. 
Therefore, the pixel size becomes finer by the reduction of $v_{x, y}$ and/or the extension of $h$.

The acquired timestamp-data are processed according to the abovementioned procedures on a data-processing computer to obtain a neutron transmission image.

\section{Detector structure and experimental apparatus}
\subsection{Detector structure}
Seven layers were deposited on a thermally oxidized Si substrate in our CBKID.
They were sequentially stacked from bottom to top as follows: (1) a 625-$\mu$m thick silicon substrate, (2) a 300-nm-thick SiO$_2$ layer, (3) a 300-nm thick superconducting Nb ground plane, (4) a 350-nm thick insulating SiO$_2$ layer, (5) a superconducting 40-nm-thick Nb $Y$ meander line, (6) a 50-nm thick insulating SiO$_2$ layer, (7) a superconducting 40-nm-thick Nb $X$ meander line, (8) a 50-nm thick passivation SiO$_2$ layer, and (9) a $^{10}$B neutron capture layer.
The nuclear reaction $^{10}$B(n, $^4$He)$^7$Li mainly emitted a $^4$He ion of 1.47\,MeV and a $^7$Li ion of 0.88\,MeV. The local energy dissipation to the meander line provided by each projectile was used to create a hot spot on the Nb meander lines.
In this detector, the $^{10}$B layer was made by painting a mixture solution of GE7031 varnish and $^{10}$B powder with a brush.
This method intends to achieve sufficient thickness compared to the ion ranges, but causes the influence of inhomogeneity in the $^{10}$B density in the conversion layer because of the segregation of the GE7031 varnish upon drying.
All turning points of the meander lines were rounded. The width was kept constant to ensure smooth propagations without the reflection of the electromagnetic waves along the whole meander line.
Moreover, the line width was gradually tapered from the end of a meander line to the electrode pad to prevent the signal reflection caused by the sudden change in impedance.
The $X$ and $Y$ meander lines of 0.9\,$\mu$m width and 15.1\,mm segment length were folded 10,000 times with 0.6\,$\mu$m spacing.
The repetition period $p$ was 1.5\,$\mu$m, and the total length of the meander line $l$ reached 151\,m.
The Nb meander line with two end electrodes was fabricated in the Clean Room for Analog-Digital Superconductivity (CRAVITY) at the National Institute of Advanced Industrial Science and Technology (AIST).
Compared with the previously reported detector \cite{CB-KID_4}, we extended the segment length to be 1.5 times longer, and the pitch width was reduced to 3/4.
The extension of the segment length tended not only to increase in the sensor detection area, but also to improve the spatial resolution with the assistance of the segment pitch refinement.
Although the ultimate pixel size of our detector may be defined by the repetition period of the meander line, the actual pixel size was an integral multiple of the repetition period because of the limitation of the temporal resolution in the readout circuit.

\subsection{Experimental apparatus}
The experimental apparatus is schematically shown in Fig.~\ref{CBKID}~(b).
The DC bias currents $I^x_b$ and $I^y_b$ were applied by two constant voltage sources through the 3\,k$\Omega$ resistors for both meander lines.
The signals from Ch1, Ch2, Ch3, and Ch4 were amplified via a differential ultralow-noise amplifier (SA-430 F5 by NF Corporation), while the negative signals from Ch1 and Ch3 were inverted. A readout board (Kalliope-DC) and a 2.5 GHz sampling digital oscilloscope (Teledyne LeCroy HDO4104-MS) simultaneously received positive signals because the positive thresholds for the counting signals in the Kalliope-DC board were configured for convenience.
The Kalliope-DC readout circuit had a 1 ns-sampling multichannel (16\,Ch\,$\times$\,2) time-to-digital converter (TDC), which was originally developed by Kojima {\it et al} for the muon-spin relaxation ($\mu$SR) measurements at the J-PARC facility \cite{Kalliope}.
CBKID and test samples were cooled down to low temperatures below $T_{\rm c}$ using a Gifford-McMahon (GM) cryocooler.
The detector temperature was monitored using a Cernox thermometer and controlled using a heater placed near the detector.
The neutron beam was irradiated to the detector from the silicon substrate side through the test samples placed at a distance of 0.8\,mm from the detector and cooled down with the detector.
Further, neutron-irradiation experiments were performed with pulsed neutrons having the collimator ratio of $L/D=14\,{\rm m}/0.10\,{\rm m}=140$ at BL10 of the Material and Life Science Experimental Facility (MLF) of J-PARC \cite{BL10}.
The neutron energy is proportional to square of velocity. Therefore, the measurement of the neutron flight time to travel the known distance provides the neutron energy. This is so-called TOF method.   
The energy resolution was achieved using the TOF method through the 14-m flight path with 33\,$\mu$s full width at half maximum (FWHM) at 10\,meV.

\section{Results and discussion}
\subsection{Signals by the neutron reaction in the CBKID}
\begin{figure}
\begin{center}
\includegraphics[scale=0.5]{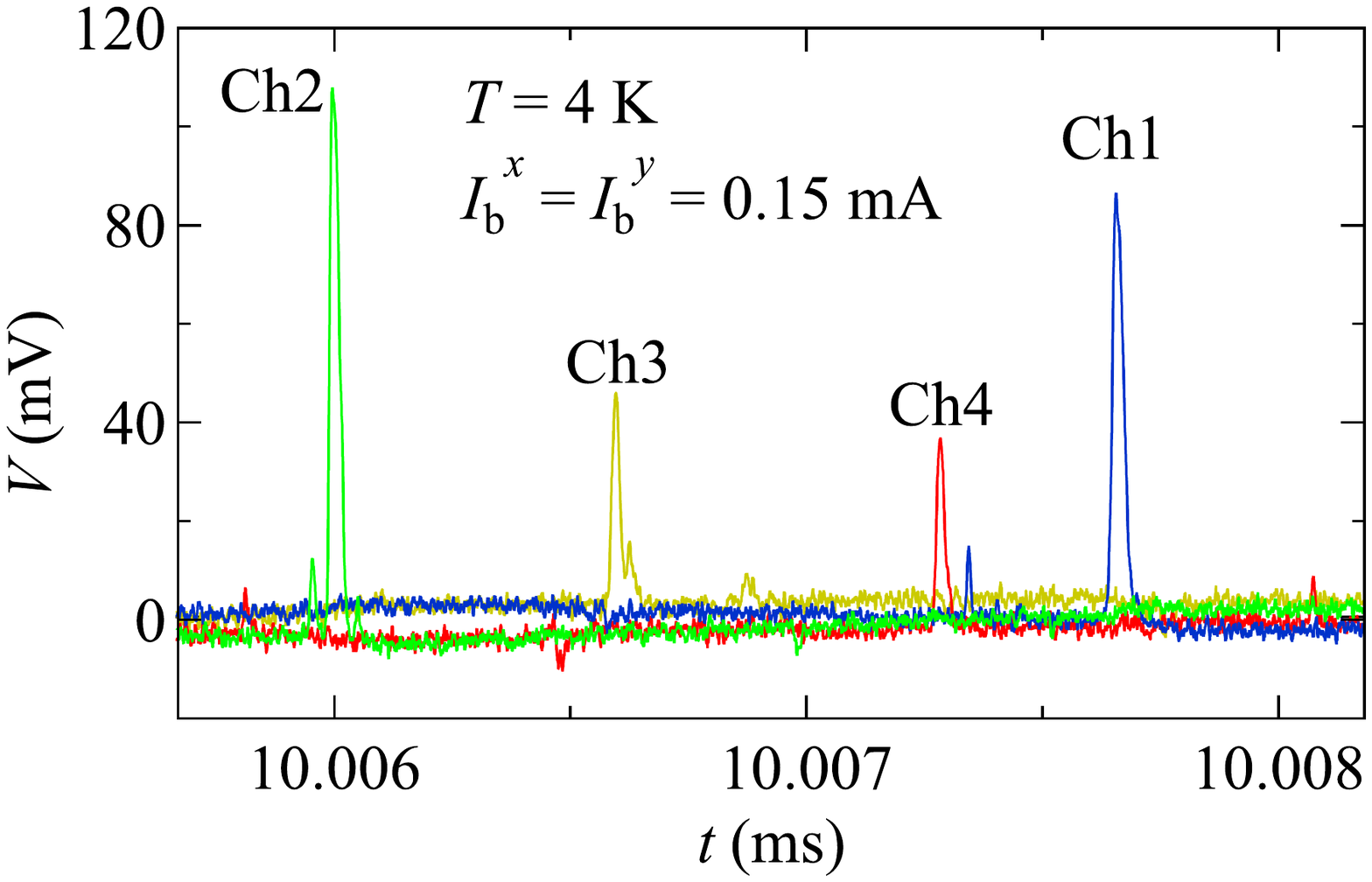}
\end{center}
\caption{Typical signal quartet from Ch1, Ch2, Ch3, and Ch4 at $T=4.0\,$K with a bias current of $I_{\rm b}^x = I_{\rm b}^y$ = 0.15\,mA.
Notably, the negative signals from Ch1 and Ch3 are inverted by the low-noise differential amplifier.
Pulsed neutrons are generated at $t=0\,$ms of the horizontal axis.
The four signals are simultaneously generated by a single nuclear reaction event.
We can find the time and coordinate of the event occurrence from the timestamps of these signals.
}
\label{Signal}
\end{figure}

Figure~\ref{Signal} shows a typical signal quartet measured using the oscilloscope.
Ch1 and Ch2 corresponded to both ends of the $Y$ meander line, whereas Ch3 and Ch4 corresponded to both ends of the $X$ meander line.
Notably, the negative signals from Ch1 and Ch3 were inverted to positive signals using a differential amplifier.
In conclusion, these four different signals arose from the single-neutron reaction event at the hot spot.
From Eqs. (\ref{eq:X}) and (\ref{eq:Y}), using the time at which these four signals were detected, the position of the hot spot created by the neutron reaction was specified as a 2D coordinate.
The signal widths of these signals were approximately 40\,ns, showing a sharp reaction.
These signal widths and signal quartet selecting procedure, which are the characteristics of the CBKID, enabled a high-speed measurement and energy dispersive neutron imaging with the combination of the TOF technique.
We estimated the detection rate tolerance to be as high as a few tens of MHz for the theoretical limit because the CBKID can discriminate multi-hit events in contrast with other techniques.
As a matter of fact, the detection rate of 0.2\,MHz can be read using our current system.

\subsection{Direct beam measurement and image processing}\label{background}
\begin{figure}
\begin{center}
\includegraphics[scale=0.5]{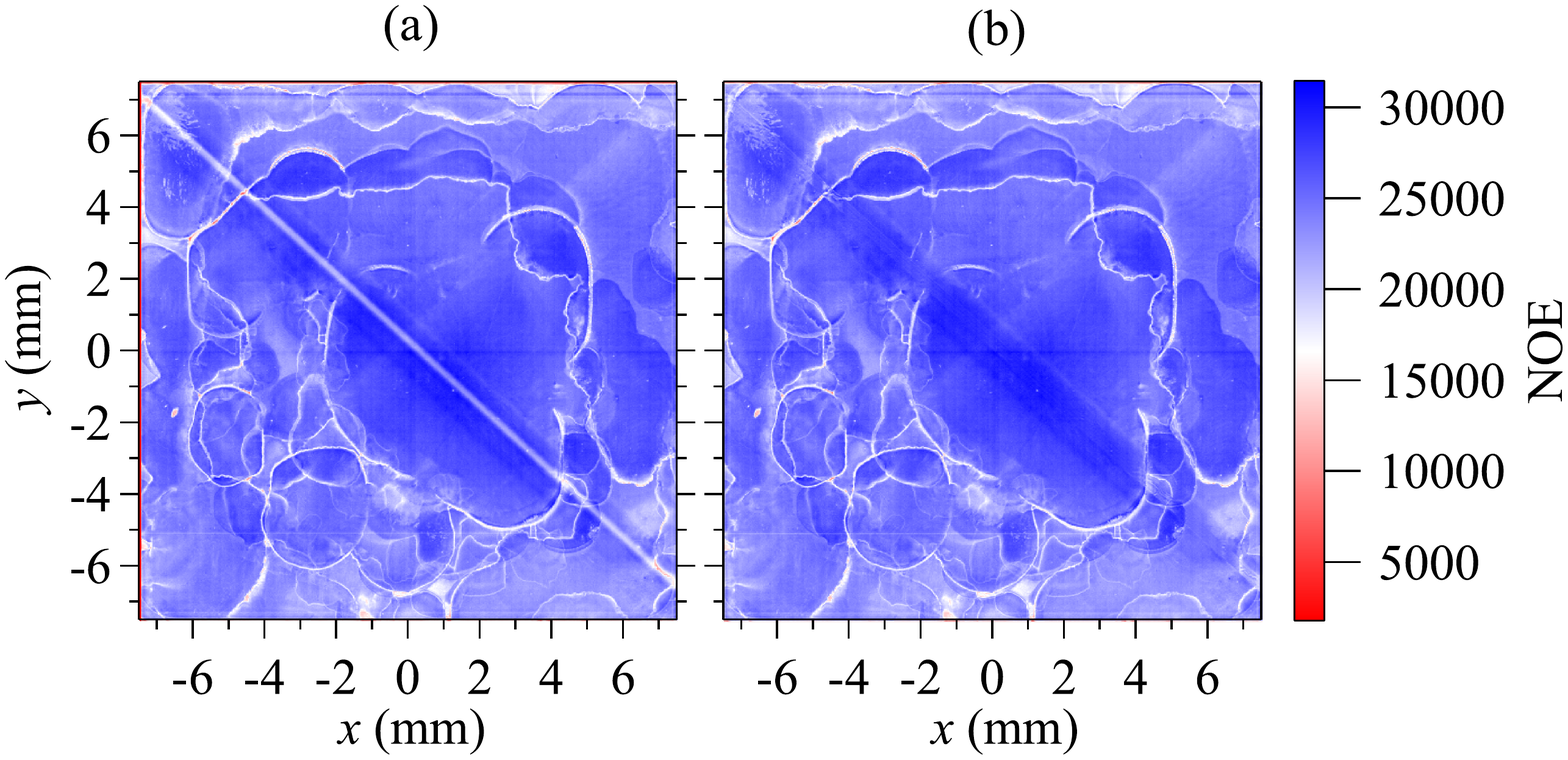}
\end{center}
\caption{(a) Direct beam neutron image using the CBKID without mounting the test samples with a wavelength from 0.052 to 1.130\,nm.
(b) Direct beam neutron image after removing the diagonal line of Fig.~\ref{DirectBeam}~(a) by image processing.
}
\label{DirectBeam}
\end{figure}

We showed the neutron transmission image herein using the CBKID without any test samples.
The color scale indicates the number of events (NOE).
The image was obtained by summing 17.9\,h under the condition of $T=4.0\,$K with $I_{\rm b}^x = I_{\rm b}^y = $0.15\,mA and 395\,kW beam power.
Figure~\ref{DirectBeam}~(a) shows the image with an incident neutron wavelength $\lambda$ ranging from 0.052 to 1.13\,nm.
The NOEs from $10\times10$ pixels were combined in Figs.~\ref{DirectBeam}~(a) and (b) to obtain a high-contrast image.

The neutron conversion of $^{10}$B layer was not homogeneous enough in the present CBKID sensor, as evidenced by the irregular curves seen in Fig.~\ref{DirectBeam}~(a).
Additionally, a white diagonal line from the upper left to lower right can be seen.
We considered this line to be caused by the signal leaks between $X$ and $Y$ meander lines.
The actual signal weakens if it is opposite in polarity with leak signal and they merge at a leak point.
Assuming that the signals arising from the neutron reaction at $(n_x, n_y)$ weakens each other at a leak point $(n_x^l, n_y^l)$, the relation between $(n_x, n_y)$ and $(n_x^l, n_y^l)$ satisfies the following equation:
\begin{eqnarray}
\label{eq:line1}
\frac{n_x^l-n_x}{v_x}=\frac{n_y-n_y^l}{v_y},
\end{eqnarray}
where $v_x$ and $v_y$ are the propagation velocities for the $X$ and $Y$ meander lines, respectively.
We can obtain the linear function from Eq.~(\ref{eq:line1}) as follows:
\begin{eqnarray}
\label{eq:line2}
n_y=-\frac{v_y}{v_x}n_x+n_y^l+\frac{v_y}{v_x}n_x^l.
\end{eqnarray}
From Eq.~(\ref{eq:line2}), the diagonal line can be predicted as a linear function with a slope of $-v_y/v_x = -5.74672\times 10^7/6.29011\times 10^7 = -0.913612$.
The diagonal line slope was obtained as $-0.9135$ from Fig.~\ref{DirectBeam}~(a), and was in a good agreement with our consideration.
The neutron imaging of the test samples was performed before the direct beam measurement and had no diagonal line, implying that the diagonal line appeared because of the aging degradation of the detector.
We tried to remove the diagonal line via imaging processing to proceed with the background correction (Fig.~\ref{DirectBeam}~(b)).

\subsection{Neutron imaging of the test samples}
\begin{figure}
\begin{center}
\includegraphics[scale=0.5]{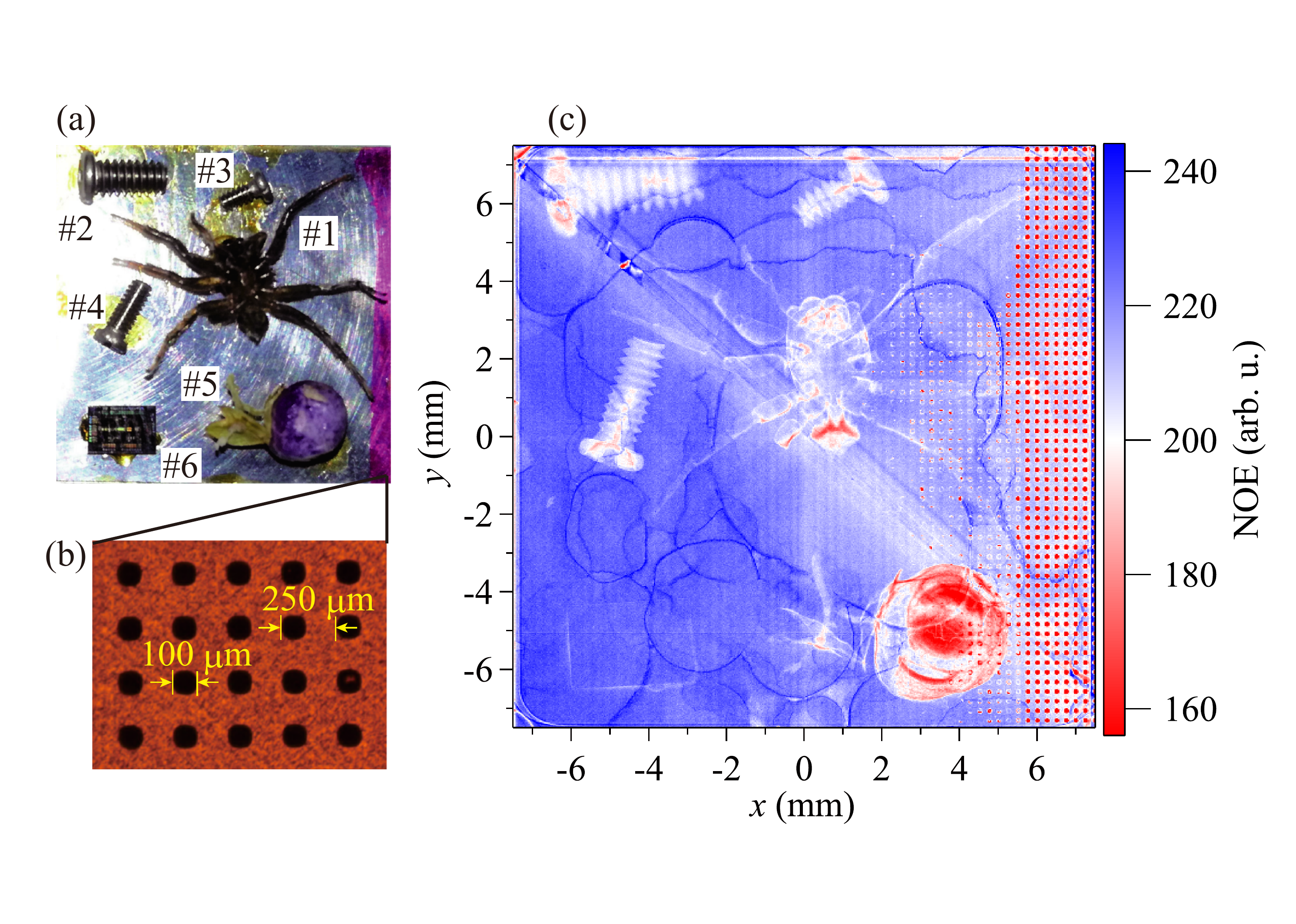}
\end{center}
\caption{(a)Photograph of the imaging samples of (\#1) a spider, (\#2) a titanium screw, (\#3) a screw of stainless-steel, (\#4) a screw of stainless-steel, (\#5) a Japanese beautyberry (plant), and (\#6) a circuit board.
(b) Optical photograph of a neutron absorber of $^{10}$B-dots as a test pattern comprising a 50 $\mu$m-thick stainless-steel mesh.
Each hole is tightly filled by very fine $^{10}$B particles.
(c) Neutron transmission image after correcting for background by dividing the neutron image with test samples by the image of Fig.~\ref{DirectBeam}~(b).
}
\label{SampleImage}
\end{figure}

Figure~\ref{SampleImage}~(a) shows a photograph of the test objects, namely (\#1) a spider, (\#2) a titanium screw, (\#3) a screw of stainless-steel, (\#4) a screw of stainless-steel, (\#5) a Japanese beautyberry(plant), and (\#6) a circuit board.
In addition, we superimposed well-shaped $^{10}$B-dots as a neutron absorber to test the spatial resolution, as shown in Fig.~\ref{SampleImage}~(b).
The test absorber comprised a 50-$\mu$m thick stainless-steel mesh (15\,$\times$\,15\,mm$^2$ in size), wherein 100\,$\times$\,100\,$\mu$m$^2$ square holes were arrayed in a square lattice (lattice constant: 250\,$\mu$m).
Each hole was tightly filled by very fine $^{10}$B particles. The stainless-steel mesh was fabricated using the wet etching technique; hence, the corners and edges of the square hole were somewhat rounded [refer the optical photograph shown in Fig.~\ref{SampleImage}~(b)].
The measurement was performed for 104.9\,h under the conditions of a bias current $I_{\rm b}=0.15\,$mA, a temperature $T$=4.0\,K, and a 304\,kW beam power.

Figure~\ref{SampleImage}~(c) shows the neutron transmission image with incident neutron wavelength $\lambda$ ranging from 0.052 to 1.13\,nm after correcting for background by dividing the neutron image with the test samples by the direct beam image of Fig.~\ref{DirectBeam}~(b).
Notably, the NOEs from $10\times10$ pixels were combined.
The plant fruit, three screws, spider, and $^{10}$B-dot pattern could be confirmed, demonstrating the capability of neutron imaging for organic and metal samples by the CBKID.
Moreover, an internal structure can be seen inside the two stainless-steel screws. 
Such a structure was not seen in the titanium screw.
Additionally, the difference between pulp and seed part of the internal structure of the berry can be seen (\#6).
Irregular curves still remained visible. 
As mentioned earlier, we succeeded in imaging the test objects of interests over the 15\,$\times$\,15\,mm$^2$ size using the CBKID.

\subsection{Spatial resolution}
\begin{figure}
\begin{center}
\includegraphics[scale=0.5]{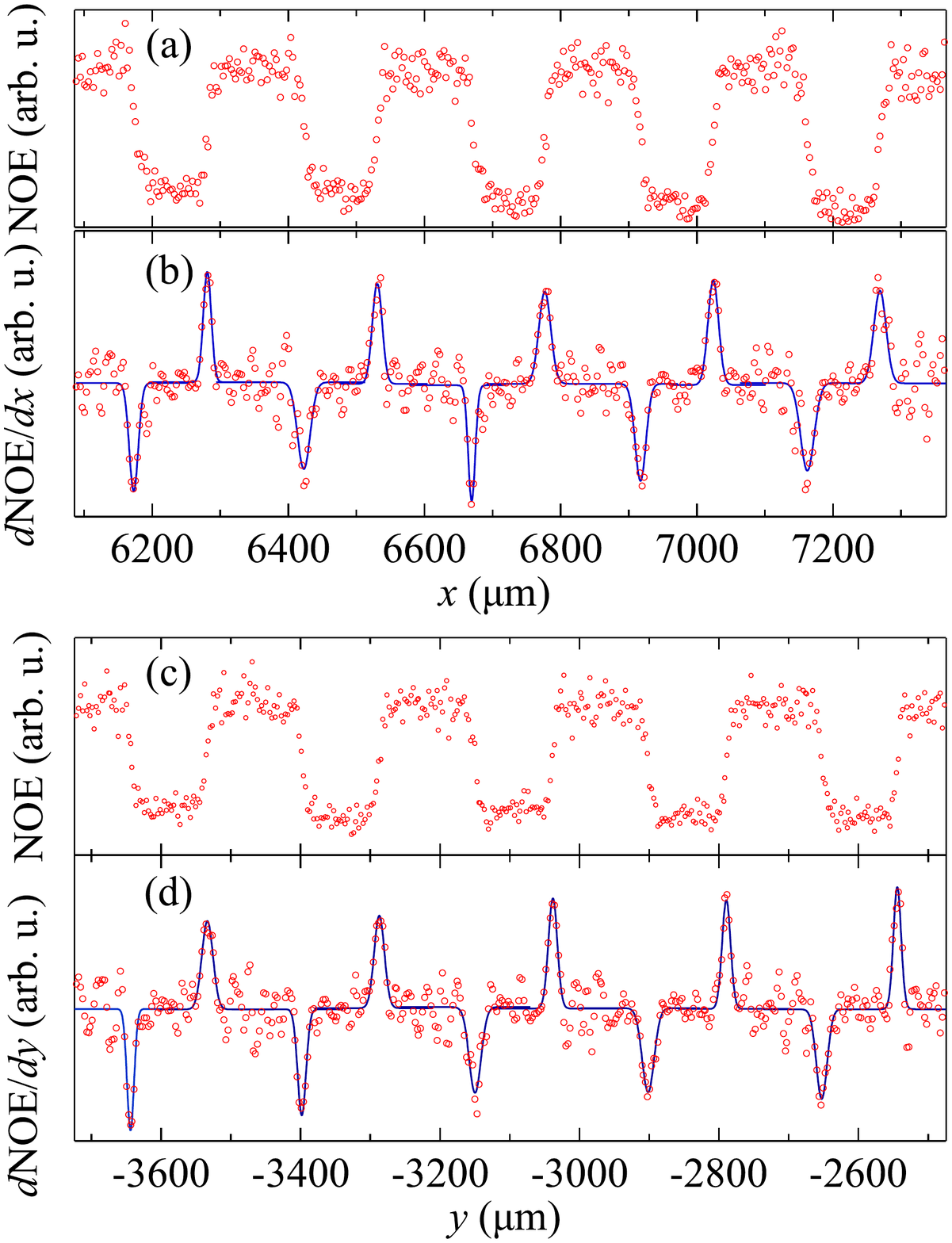}
\end{center}
\caption{(a) Number of events (NOE) on the $^{10}$B-dots boundary in the $X$-direction along the line $y=1885.5\,\mu$m.
(b) Numerical differentiation of the NOE as a function of $y$ with the least-squares fit line to a Gaussian function.
(c) NOE on the $^{10}$B-dots boundary in the $Y$-direction along line $x=6229.5\,\mu$m.
(d) Numerical differentiation of the NOE as a function of $y$ with the least-squares fit line to a Gaussian function.}
\label{LineProf}
\end{figure}

We discussed herein the spatial resolution of the CBKID using the $^{10}$B-dot pattern embedded in the test sample.
Figures~\ref{LineProf}~(a), (b), (c), and (d) show the typical line profiles along the (a) $X$ and (c) $Y$-directions and the corresponding differentiations along the (b) $X$ and (d) $Y$-directions with minimum pixel sizes of 3 or 4.5\,$\mu$m and 1.5 or 3\,$\mu$m for $X$ and $Y$-directions, respectively.
The Gaussian fitting results for the differentiations are depicted by the solid lines in Figs.~\ref{LineProf}~(b) and (d).
We examined the Gaussian fitting for differentiations in all the clear dot patterns (480\,points) appearing in Fig.~\ref{SampleImage}~(c) and obtained the averages of the FWHM as 19.2\,$\mu$m and 16.2\,$\mu$m for the $X$ and $Y$-directions, respectively.
The spatial resolution in the $Y$-direction was better than that in the $X$-direction because $v_y$ was slower than $v_x$.
The incompleteness of the edge sharpness of the hole in the stainless-steel mesh and the incomplete filling of $^{10}$B-dot particles into the hole may partially affect the results of the spatial resolution of interest.
As expected, the present spatial resolutions were improved compared to those in a previous report\cite{CB-KID_4}, in which the spatial resolutions were evaluated using a similar test pattern.

\subsection{Material analysis using the Bragg edge}
\begin{figure}
\begin{center}
\includegraphics[scale=0.5]{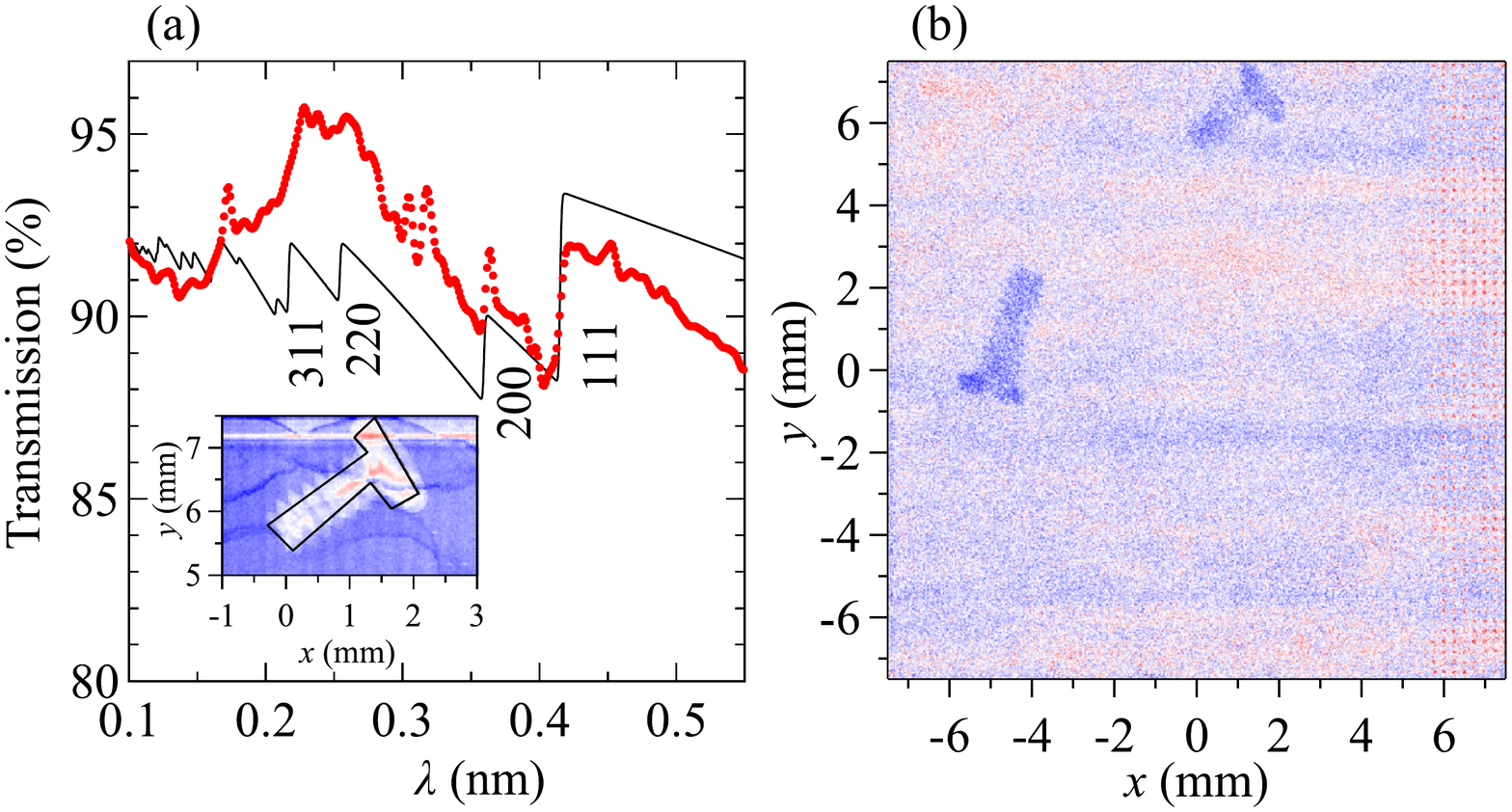}
\end{center}
\caption{(a) Neutron transmission of the stainless-steel screw (sample \#3) in the area surrounded by a solid line on an image (see inset), together with a simulation curve with the Miller indices. 
(b) Stainless-steel enhanced image obtained by dividing the neutron transmission image with a wavelength shorter than 111 Bragg edge by that with a wavelength longer than 111 Bragg edge.
}
\label{BraggEdge}
\end{figure}

The combination of the high temporal resolution in CBKID and TOF method in a pulse neutron source allowed us to demonstrate wavelength (energy) selective neutron imaging.
Figure~\ref{BraggEdge}~(a) shows the neutron transmission of the stainless-steel screw (sample \#3). 
Notably, the metal mesh of the stainless-steel was not attached during the direct beam measurements; however, it was installed during neutron imaging measurements of the test samples.
The 111 stainless-steel Bragg edge was clearly observed, as shown in Fig.~\ref{BraggEdge}~(a).
A finite, but distinct 200 Bragg edge also arose.
Figure~\ref{BraggEdge}~(b) shows the image ratio of the neutron transmission image with a wavelength shorter than 111 Bragg edge (0.390(1)\,nm$<\lambda<$0.401(1)\,nm) and that with a wavelength longer than 111 Bragg edge (0.423(1)\,nm$<\lambda<$0.434(2)\,nm). 
As shown in Fig.~\ref{BraggEdge}~(b), only stainless-steel screws were clearly observed. The other samples disappeared by division.

\section {Summary}
In this study, we demonstrated higher spatial resolution neutron imaging compare with the previous report \cite{CB-KID_4} and Bragg edge analysis by delay-line CBKID systems.
Accordingly, we succeeded in fabricating 0.9-$\mu$m width strip lines without any disconnection even in a 151\,m length.
This improvement of the CBKIDs brought a high spatial resolution down to 16.2\,$\mu$m.
Test samples with various shapes and materials were clearly observed over the whole sensor area of 15\,$\times$\,15\,mm$^2$.
In addition, our detector has a temporal resolution in combination with the Kalliope-DC readout circuit with 1 ns resolution TDC for high-speed data acquisition.
By combining with the TOF method, the delay-line CBKID became capable of wavelength (energy)-selective neutron imaging and Bragg edge analysis, as demonstrated for stainless-steel.
A further improvement of the fabrication method of the homogeneous $^{10}$B-conversion layer is required. Nonetheless, the CBKID has potential as a unique neutron imager.

\ack
This work is partially supported by a Grant-in-Aid for Scientfic Research (Grant Nos. JP16H02450, JP19K03751) and from JSPS.
The neutron-irradiation experiments at the Materials and Life Science Experimental Facility (MLF) of the J-PARC are conducted under the support of MLF programs (Proposals Nos. 2015A0129, 2016B0112, 2017A0011, 2017B0014, 2018A0109, 2015P0301, 2016P0301, 2018P0201).
Development of the Kalliope TDC and readout electronics/software is conducted under the collaboration of KEK Open Source Consortium of Instrumentation (Open-IT).

\clearpage

\clearpage

\clearpage

\clearpage

\clearpage

\clearpage

\clearpage

\end{document}